\documentclass[prl,twocolumn,superscriptaddress,letter ]{revtex4}
\usepackage{graphicx}
\usepackage{wasysym}
\def\HOmI{\mbox{HOMO-1}}
\def\HOmII{\mbox{HOMO-2}}
\def\intensity[#1][#2]{\mbox{$#1\times10^{#2}$~W/cm$^2$}}
\def\h2o{H$_2$O}
\def\hs{\hspace{1.5pt}}
\def\PULSE{
    Stanford PULSE Institute, SLAC National Accelerator Lab,
    2575 Sand Hill Road, Menlo Park CA 94025 and
    Departments of Physics and Applied Physics, Stanford University, Stanford CA 94305
}

\def\Berlin{
     AG Moderne Optik,
     Institut f\"ur Physik,
     Humboldt-Universit\"at zu Berlin,
     Newtonstr. 15, D\,-\,12\,489 Berlin, Germany
}
\def\Trieste{
      Dipartimento di Scienze Chimiche,
      Universit{\`a} di Trieste,
      Via~L.~Giorgieri 1,
      I\,-\,34127 Trieste, Italy
}

\begin{document}
\title{Strong field ionization to multiple electronic states in water}

\author{Joe P. Farrell} \affiliation{\PULSE}
\author{Simon Petretti} \affiliation{\Berlin}
\author{Johann F\"orster} \affiliation{\Berlin}
\author{Brian K. McFarland} \affiliation{\PULSE}
\author{Limor S. Spector} \affiliation{\PULSE}
\author{Yulian V. Vanne} \affiliation{\Berlin}
\author{Piero Decleva} \affiliation{\Trieste}
\author{Philip H. Bucksbaum} \affiliation{\PULSE}
\author{Alejandro Saenz} \affiliation{\Berlin}
\author{Markus G\"uhr} \affiliation{\PULSE}

\begin{abstract}
High harmonic spectra show that laser-induced strong field ionization of water has a significant contribution from an inner-valence orbital. Our experiment uses the ratio of H$_2$O and D$_2$O high harmonic yields to isolate the characteristic nuclear motion of the molecular ionic states. The nuclear motion initiated via ionization of the highest occupied molecular orbital (HOMO) is small and is expected to lead to similar harmonic yields for the two isotopes. In contrast, ionization of the second least bound orbital (\HOmI) exhibits itself via a strong bending motion which creates a significant isotope effect. We elaborate on this interpretation by simulating strong field ionization and high harmonic generation from the water isotopes using the time-dependent Schr\"odinger equation. We expect that this isotope marking scheme for probing excited ionic states in strong field processes can be generalized to other molecules.
\end{abstract}
\maketitle

The superposition of several ionic states resulting from strong field ionization of molecules has been observed via high harmonic generation (HHG). Tunnel ionization preferentially ionizes the HOMO. This leaves the molecule in the ground ionic state. Rotational or vibrational pre-excitation is typically used  to force ionization from inner valence orbitals \cite{McFarland_2008, Smirnova_2009b, Li_2008}. The population of ionic states is then probed by the recombination step of HHG \cite{Kulander_1993, Corkum_1993, Schafer_1993}. In this contribution we show that strong field ionization of water can populate several ionic states \emph{without} vibrational or rotational pre-excitation. This is done with an isotropic, unexcited ensemble by using isotope marked molecules. The excitation of lower lying valence states triggers an electronic wave packet with femtosecond to attosecond time scale dynamics \cite{Loh_2007}, which was demonstrated recently for the case of  fine-split ionic states of Krypton \cite{Goulielmakis_2010}.

\begin{figure}[b] \centering
\includegraphics[width=3in,]{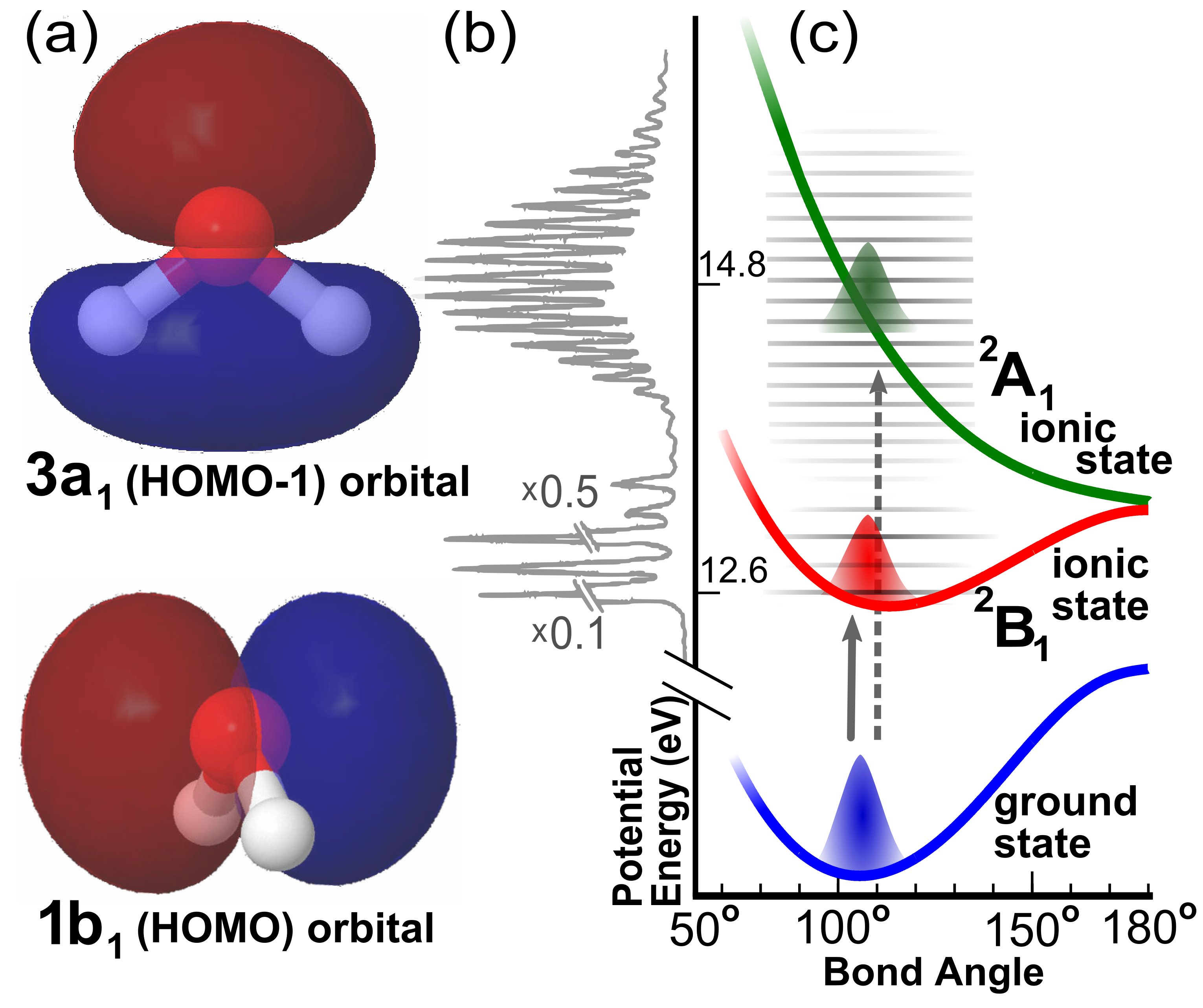}
\caption {(a) Constant amplitude contours of the HOMO ($1b_1$) and \HOmI ($3a_1$) orbitals. (b) Photoelectron spectrum of H$_2$O reproduced from \cite{wat:karl75}. The transition to the ground ionic state ($^{2}$B$_{1}$) occurs near 12.6~eV and populates only a few vibrational lines. In contrast, the transition to the first excited ionic state ($^{2}$A$_{1}$), which occurs around 14.8~eV, populates many vibrational lines. (c) Potential energy curves of the neutral and first two ionic states. The minimum of the $^{2}$B$_{1}$ surface lies almost directly above that of the neutral, while the minimum of the $^{2}$A$_{1}$ surface is at 180$^{\circ}$.}
\label{diagram}
\end{figure}
We investigate ionization of the inner valence orbital by comparing the high harmonic yields from different isotopes of water. As first understood by Lein and observed by Baker \emph{et al.}, nuclear dynamics occurring between ionization and recombination are mapped onto the harmonic spectrum and can be isolated by using isotope marked molecules \cite{LeinM:Attpvd,Baker_2006}. Thus, the presence of multiple ionic states in HHG can be determined if the different ionic states launch significantly different nuclear motion.

Figure \ref{diagram} presents an intuitive picture of the theory elaborated in \cite{LeinM:Attpvd}. The ionization of the molecular orbital (either HOMO or \HOmI $\hs$ in Fig.~\ref{diagram}(a)) launches a vibrational wave packet on an ionic potential energy surface in Fig.~1(c) (red and green, respectively). The harmonic light is emitted as the returning electron recombines with the molecular ion. Photoionization and HHG recombination are inverse processes \cite{Farrell_2011}. The well known modulation of extreme ultraviolet photoionization cross sections by the Franck-Condon overlap (shown in Fig.~\ref{diagram}(b) for \h2o) then requires the HHG recombination cross section to be modulated by the overlap of the ionic state nuclear wave packet and the neutral vibrational ground state. As the ionic state nuclear wave packet moves, it loses overlap with the ground state. The nuclear wave packet for the heavy isotope is slow, thus maintaining a stronger overlap than the lighter, faster isotope. This gives a larger harmonic yield for the heavier isotope. The timescale of the nuclear motion is mapped onto the energy of different harmonics \cite{LeinM:Attpvd}. Comparison of harmonic spectra from the different isotopes isolates the effect of nuclear motion, since their electronic structure is identical.

In H$_2$O, population of different ionic states can launch significantly different vibrational motion. Ionization of the HOMO, $1b_1$, excites the molecule into the ionic ground state, $\tilde{X}{}^{2}$B$_{1}$. The resulting potential energy surface [Fig.~\ref{diagram}(c)] is very similar to that of the neutral as can be seen from the exponentially decaying Franck-Condon progression of the $\tilde{X}{}^{2}$B$_{1}$ state in the photoelectron spectrum shown in Fig.~\ref{diagram}(b). This is due to the non-bonding properties of the $1b_1$ orbital evident by its shape: the $1b_1$ is a single oxygen $p$ orbital that sticks out of the molecular plane [Fig.~\ref{diagram}(a)]. In turn, only extremely \emph{weak} nuclear dynamics are induced  \cite{wat:karl75, wat:reut86}. In contrast, ionization of the \HOmI $3a_1$, which sends the molecule into the $\tilde{A}{}^{2}$A$_{1}$ state, \emph{strongly} excites the bending mode \cite{wat:karl75, wat:reut86}. The equilibrium bond angle changes to 180$^{\circ}$ [Fig.\ref{diagram}(c)]. The $3a_1$ has an intramolecular bonding character, see Fig.~\ref{diagram}(a). The photoelectron spectrum illustrates the strong (weak) vibrational excitation upon ionization of the $3a_1$ ($1b_1$) orbital. The vertical binding energies of the $3a_1$ and $1b_1$ orbitals are 14.8 and 12.6~eV, respectively \cite{Ning_2008}.

The very weak vibrational wave packet excited via population of the ionic ground state indicates that HHG from the $1b_1$ should not display a strong isotope effect. This is confirmed by a recent simulation, which considered ionization solely from the $1b_1$ and found a negligible difference between H$_2$O and D$_2$O high harmonics for vibrationally unexcited neutral molecules \cite{sfm:falg10}.

Several experiments have shown that inner valence orbitals can contribute significantly to high harmonic yields \cite{McFarland_2008,Smirnova_2009, Smirnova_2009b, Li_2008}. We expect the inner valence $3a_1$ orbital to contribute to the high harmonics of water. The broad vibrational wave packet launched via ionization of the $3a_1$ orbital causes a rapid decrease in the overlap between the ionic and ground state nuclear wave functions. This results in a reduction of the HHG yield and significant difference between H$_2$O and D$_2$O harmonics. Water is therefore an ideal test molecule to determine inner valence contributions via nuclear motion. Harmonic generation in water has been demonstrated before in both liquid and gas phase samples \cite{DiChiara_2009, Wong_2010}.

We generate harmonics with a commercial Ti:Sapphire laser (pulse duration 30$\hs$fs, pulse energy 100$\hs$$\mu$J, central wavelength 800$\hs$nm, repetition rate 1$\hs$kHz). A 40$\hs$cm focal length mirror focuses the beam to a full width at half maximum of about {$\mathrm{60}\hs\mu$m}. The harmonics between 20 and 70~eV pass through an Al filter (thickness 100~nm) onto a toroidal flat-field grating, which disperses the radiation onto an extreme ultraviolet sensitive CCD camera. The gas cell is approximately 0.5$\hs$mm thick and has entrance/exit holes of approximate diameter 0.2$\hs$mm. Water and heavy water reservoirs kept at room temperature are connected to the cell by a fine-dosing valve to keep the cell gas pressure of about 3$\hs$mbar (number density 10$^{17}$cm$^{-3}$). The laser intensity is changed by inserting pellicle beam-splitters into the beam. An iris placed just upstream of the grating rejects the long trajectory harmonics \cite{Salieres_1995}.

\begin{figure}\centering
\includegraphics[width=3in,]{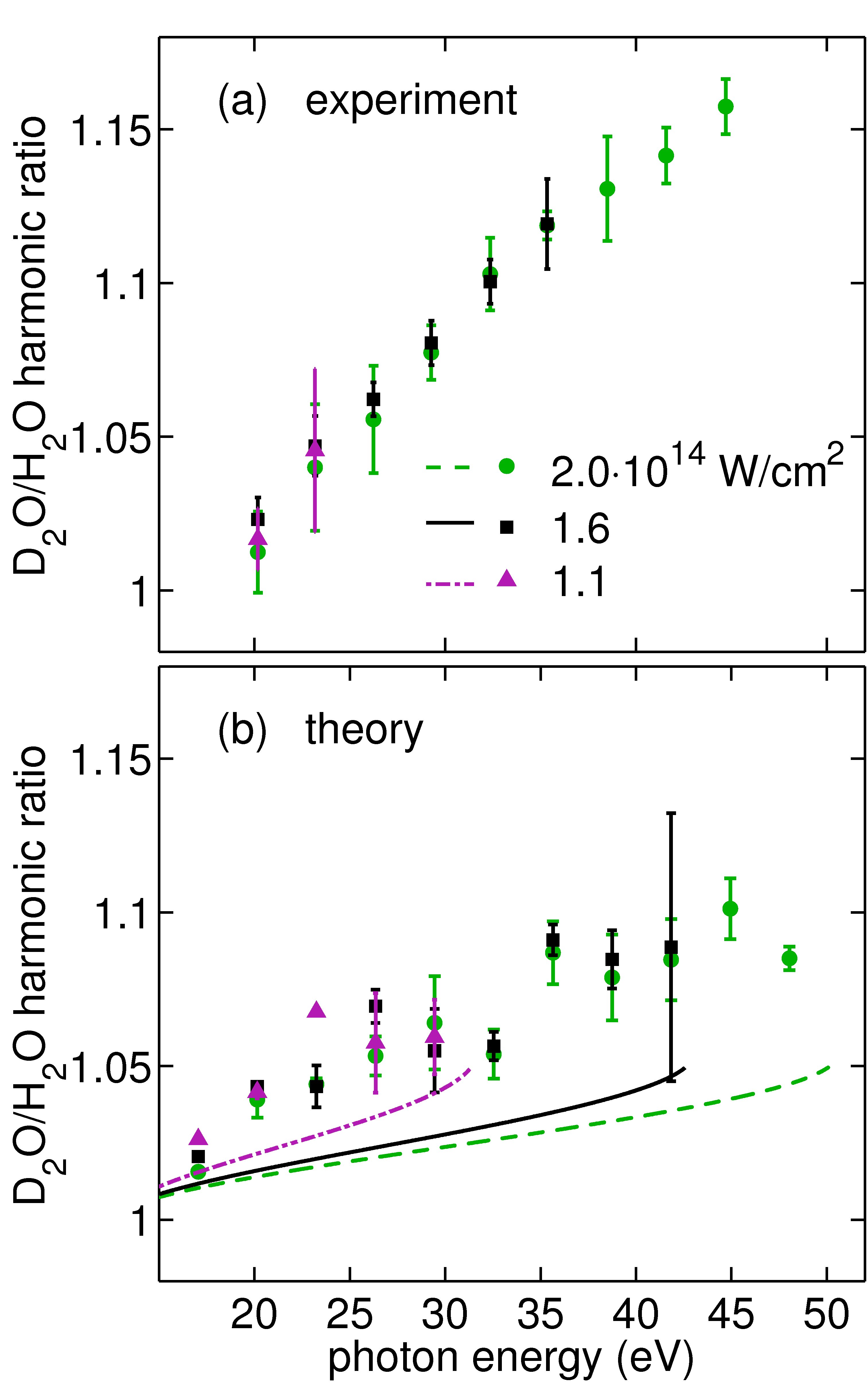}
\caption {(a) Heavy water to water harmonic yield ratios for several intensities. The error bars include the standard error of the mean of the individual datasets, but are dominated by a systematic effect characterized by taking two different sets of spectra at different times. (b) Simulated ratios. Markers represent simulations including both the $1b_1$ and $3a_1$ orbitals for several intensities. Lines represent simulations that solely include the $1b_1$. See text for description of the theoretical error bars.}\label{Spectra_and_ratios}
\end{figure}

Figure~\ref{Spectra_and_ratios}(a) shows the experimental ratios of heavy water to water spectra for several intensities spanning 1$\hs$-$\hs$2$\times$10$^{14}$W/cm$^2$. Ratios calculated using odd harmonic peak height or area had negligible differences. The error bars are dominated by a systematic effect characterized by taking two different data sets of spectra at different times. The ratio increases monotonically with photon energy for all intensities. In the range 20-50~eV, the ratio spans values from 1 to 1.15. Also, changes in intensity did not measurably affect the ratio of any given harmonic, effectively keeping the slope of the curve constant.

These results show that the D$_2$O/H$_2$O harmonic ratio gets progressively larger for longer electron travel times. Thus strong field ionization launches significant and fast nuclear motion in water. As discussed above, the photoionization studies as well as the disagreement with the $1b_1$-based simulation in \cite{sfm:falg10} suggest that ionization of the $3a_1$ is the dominant source of this motion.

To model the experiment, we extend the theoretical approach used in \cite{sfm:falg10} by taking into account multiple electronic states from different molecular orbitals. Furthermore, we solve the time-dependent Schr\"odinger equation under the single-active-electron approximation (SAE) with fixed nuclei separately for each orbital. Our approach has been used successfully for H$_2$ \cite{sfm:awas08} as well as N$_2$, O$_2$, and CO$_2$ \cite{sfm:petr10}. The resulting spectrum is then multiplied by the nuclear autocorrelation function to include the nuclear motion.

We now briefly outline this theoretical method, for details see references \cite{sfm:awas08,sfm:petr10}. Field-free Kohn-Sham orbitals are obtained using the LB94 exchange-correlation potential \cite{gen:vanl94}. They are expressed in a multi-centered basis \cite{bsp:toff02}, which consists of a set of (typically atom-centered) local spheres (bases) defined as the product of spherical harmonics and a radial part expressed in $B$-splines. Molecular symmetry (for water $C_{2v}$) is accounted for by implicitly generating equivalent local basis sets at symmetry-equivalent positions. A large central sphere positioned at the charge center of the molecule defines an additional large $B$-spline basis. It overlaps with all atomic-centered spheres and serves for an improved description of the chemical bonding and the molecular electronic continuum. The latter is discretized with a density determined by the size of the central sphere. The radius of the central sphere is set to $r_{\mathrm{max}}^0=121.5$~a.\,u., the maximum number of angular momenta used is $l_{max}=12$ and the experimentally determined equilibrium geometry ($R_{OH}$\hs=\hs0.958$\mathring{A}$,  $\varangle$HOH\hs=\hs104.5\ensuremath{^\circ} \cite{gen:herz66}) is applied. The time-dependent electronic wavefunction $|\Psi_{\alpha}(t)\rangle$ is represented as the sum of the Kohn-Sham orbitals multiplied by time-dependent coefficients. The index $\alpha$ refers to the orbital that is initially occupied. All Kohn-Sham orbitals are propagated in the velocity gauge during their interaction with a 10 cycle cos$^2$ laser pulse centered at 800~nm. The HHG spectrum for a single molecular orientation is then:
\begin{equation}
\vec{S}_\alpha(\omega) = |\int_{-\infty}^{+\infty} dt\, e^{i\omega t}\, \omega\, \vec{d}_\alpha(t) f(t)\, |^2
\end{equation}
where $\vec{d}_\alpha(t) = \langle \Psi_\alpha(t) | \vec{p} + \vec{A}(t) | \Psi_\alpha(t) \rangle$. The canonical momentum $\vec{p}$ and the vector potential $\vec{A}(t)$ are expressed in a.u. The damping function $f(t)$ reduces $\vec{d}_\alpha(t)$ at the extreme times to minimize edge effects in the Fourier transform.

Since the experimental sample is unaligned, we average simulated spectra from many different angles between the molecular axis and the laser polarization. This is performed for each orbital separately. Both coherent and incoherent averaging give essentially the same ratio D$_2$O/H$_2$O. Averaging over only the three main axes of water (including the main ionization axes for the HOMO, \HOmI{} and \HOmII{}) approximates the ratio obtained with about 80 orientations very well. Following Lein \cite{LeinM:Attpvd}, the influence of nuclear dynamics is included only by the autocorrelation function $C_{\alpha}(t)$ defined as
\begin{equation}
C_{\alpha}(t) = \langle \phi_{\alpha}(0) | \phi_{\alpha}^+(t) \rangle
\end{equation}
where $|\phi_{\alpha} \rangle$ denotes the vibrational ground state of H$_2$O and $|\phi_{\alpha}^+ \rangle$ denotes the time-dependent nuclear wave packet launched on H$_2$O$^+$. Data for the autocorrelation functions is taken from Reutt {\it et al} \cite{wat:reut86}. Finally, we form the incoherent sum of the orientation-averaged HHG contributions $S^{(D_2O/H_2O)}(\omega)$ from the HOMO $S_{1b_1}(\omega)$ and the \HOmI{} $S_{3a_1}(\omega)$ (the HHG contribution of the \HOmII{} $1b_2$ is found to be negligible) multiplied by the respective autocorrelation function
\begin{equation}
S^{(D_2O/H_2O)}(\omega) = \sum_{\alpha=1b_1, 3a_1}|C_\alpha^{(D_2O/H_2O)}(\omega)|^2 S_\alpha(\omega)
\end{equation}
where we map the time of the autocorrelation function to energy by using equation (12) in \cite{LeinM:Attpvd}. This restricts the spectrum to energies below the cutoff. The ratio D$_2$O/H$_2$O is then obtained by $S^{(D_2O)}(\omega)/S^{(H_2O)}(\omega)$.

Uncertainty in the theoretical ratios was determined by varying several parameters. First, three different damping functions were used. In each case the high harmonic spectra are additionally convolved with a Gaussian distribution function for four different variances chosen to maintain the basic peak structure of the spectra. Finally, the ratios using both peak heights and areas were computed for each case.

Figure \ref{Spectra_and_ratios}(b) shows simulations using the same intensities as the experiment. In agreement with \cite{sfm:falg10}, the isotope effect calculated using only the 1b$_1$ orbital (shown with lines) is very small in comparison to the measurement. The two orbital simulation (shown with markers) gives a much larger effect, confirming our qualitative estimate that the $3a_1$ plays an important role. For example, at 45~eV, the ratio predicted by the two orbital simulation, approximately 1.10, is much closer to the measured value of 1.15 than that of our single orbital simulation, 1.04.

The intensity dependence of the harmonic spectra is also telling. As stated above, the experimental slope had no discernible dependence on intensity. In contrast, increasing the intensity in the single orbital simulation decreases the isotope ratio for each harmonic, effectively reducing the slope of the curve. This is expected from the single orbital picture due to the one-to-one mapping between travel time and isotope ratio. In agreement with the measurement, the two orbital simulation does not exhibit a discernible change in slope with intensity, even though the intensity range used is extremely large.

\begin{figure}\centering
\includegraphics[width=3.5in,]{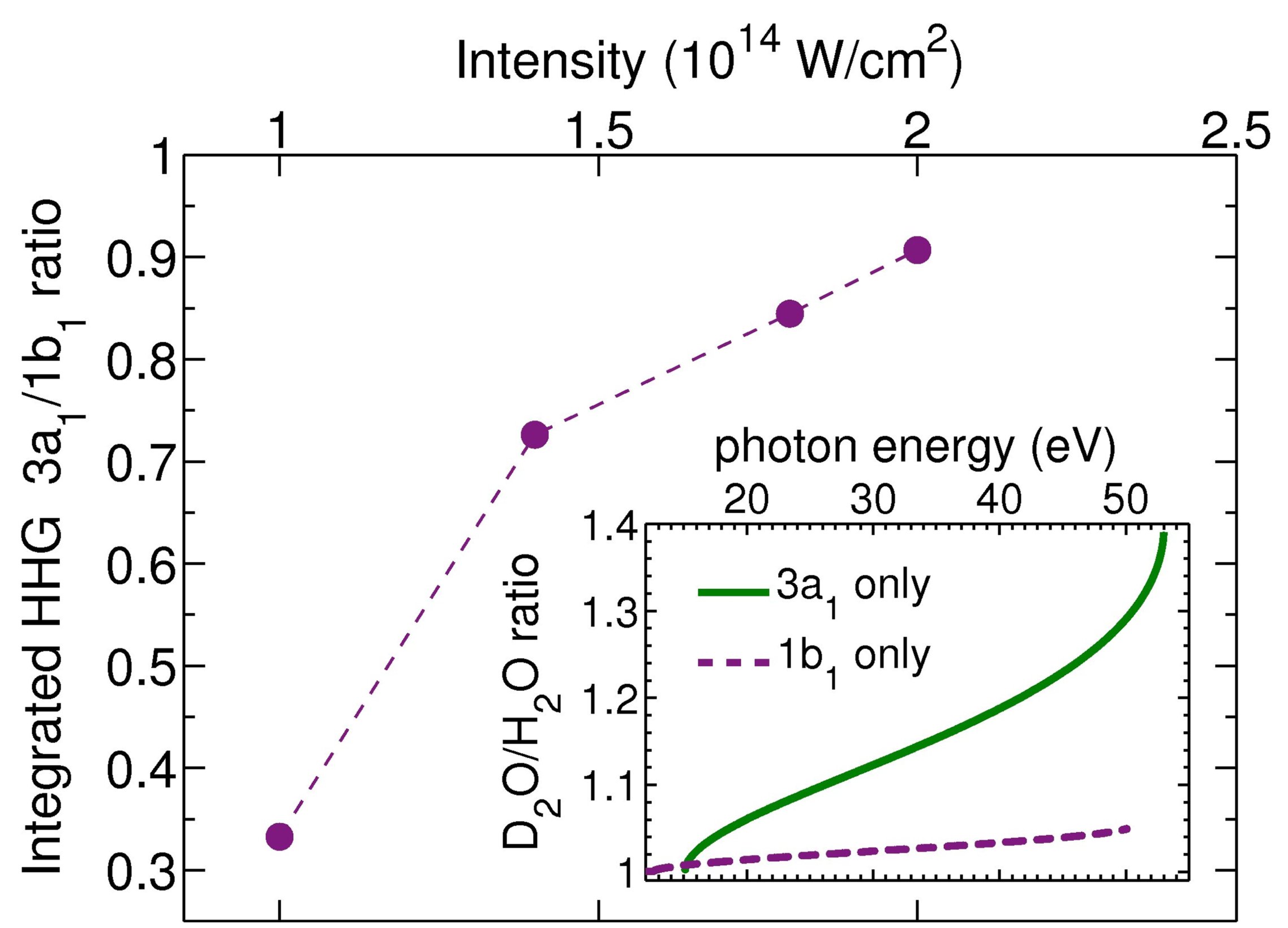}
\caption {Ratio of the calculated total harmonic yields from the $3a_1$ and $1b_1$ orbitals of H$_2$O as a function of intensity. Inset shows calculated D$_2$O/H$_2$O harmonic ratios of the $3a_1$ and $1b_1$ orbitals separately for an intensity of $2 \times 10^{14}W/cm^2$.}\label{angular_saturation}
\end{figure}

Examination of the simulations suggests that the independence of the slope on intensity stems from an angular ionization saturation effect. Prior to the laser pulse, the angular probability distribution is isotropic. For a given molecule, the primary ionization direction of the $1b_1$ (perpendicular to the molecular plane) or the $3a_1$ (along the principle axis) are both equally likely to point along the polarization direction of the laser. As the intensity is increased, ionization causes the angular distribution to be depleted along the weakly bound $1b_1$ ionization direction after several cycles of the pulse. The relative weight of HHG from the $3a_1$ then increases with intensity (see Fig. \ref{angular_saturation}). The HHG spectrum of the $3a_1$ by itself provides an estimate of the upper bound for the isotope effect and is shown in the inset in Fig. \ref{angular_saturation}. A maximum D$_2$O/H$_2$O ratio of approximately 1.4 results in this case.

We have shown, using water, that strong field ionization populates an excited electronic state without vibrational/rotational pre-excitation. This is done with an isotropic ensemble by using isotope marked molecules. We modeled the multi-orbital system by employing a well tested SAE-TDSE simulation framework and show that both the $1b_1$ and $3a_1$ orbitals contribute. The simulation predicts a magnitude of the isotope effect somewhat smaller than is measured but exhibits the same characteristic intensity dependence. The single orbital simulation predicts a much smaller effect and a different intensity dependence than is measured. The simulation suggests that the measured independence of the ratio on intensity  is due to angular ionization saturation, which increases the relative contribution of the lower valence orbital at high intensities.

These results provide a new tool for assessing contributions of different orbitals from strong field ionization. In addition, we show that ionization from inner valence orbitals can also excite nuclear dynamics. This extends the technique used in \cite{LeinM:Attpvd, Baker_2006, sfm:patc09} to a larger range of molecules. Possible improvements of the theoretical model include, e.\,g., an incorporation of the relative phase between the two orbitals \cite{Smirnova_2009b} and geometry-dependent ionization yields \cite{Saenz_2000}.

\begin{acknowledgments}
We thank the Department of Energy, Office of Basic Energy Science Science, and the
Stanford University Dean of Research for support through the Stanford PULSE Institute.
We thank COST {\it CM0702} for financial support. S.P. and A.S. acknowledge financial support from the {\it Stifterverband f\"ur die Deutsche Wissenschaft}, the {\it Fonds der Chemischen Industrie}, and {\it Deutsche Forschungsgemeinschaft} within SFB\,450. L. S. S. acknowledges financial support from a NDSEG fellowship. P.D. acknowledges financial support from CNR-INFM Democritos and INSTM Crimson. This work was supported in parts by the National Science Foundation under Grant No.\ NSF PHY05-51164.
\end{acknowledgments}


\begin{thebibliography}{10}

\bibitem{McFarland_2008}
McFarland, B.~K., Farrell, J.~P., Bucksbaum, P.~H., and G\"uhr, M.
\newblock {\em Science}{ \bf 322}, 1232 (2008).

\bibitem{Smirnova_2009b}
Smirnova, O., Mairesse, Y., Patchkovskii, S., Dudovich, N., Villeneuve, D.,
  Corkum, P., and Ivanov, M.~Y.
\newblock {\em Nature}{ \bf 460}, 972 (2009).

\bibitem{Li_2008}
Li, W., Zhou, X., Lock, R., Patchkovskii, S., Stolow, A., Kapteyn, H., and
  Murnane, M.
\newblock {\em Science}{ \bf 322}, 1207 (2008).

\bibitem{Kulander_1993}
Kulander, K.~C., Schafer, K.~J., and Krause, J.~L.
\newblock {\em Laser Physics}{ \bf 3}, 359 (1993).

\bibitem{Corkum_1993}
Corkum, P.~B.
\newblock {\em Phys. Rev. Lett.}{ \bf 71}, 1994--1997 (1993).

\bibitem{Schafer_1993}
Schafer, K.~J., Yang, B., DiMauro, L.~F., and Kulander, K.~C.
\newblock {\em Phys. Rev. Lett.}{ \bf 70}, 1599 (1993).

\bibitem{Loh_2007}
Loh, Z.-H., Khalil, M., Correa, R.~E., Santra, R., Buth, C., and Leone, S.~R.
\newblock {\em Phys. Rev. Lett.}{ \bf 98}, 143601 (2007).

\bibitem{Goulielmakis_2010}
Goulielmakis, E., Loh, Z.-H., Wirth, A., Santra, R., Rohringer, N., Yakovlev,
  V.~S., Zherebtsov, S., Pfeifer, T., Azzeer, A.~M., Kling, M.~F., Leone,
  S.~R., and Krausz, F.
\newblock {\em {Nature}}{ \bf {466}}, {739} ({2010}).

\bibitem{LeinM:Attpvd}
Lein, M.
\newblock {\em Phys. Rev. Lett.}{ \bf 94}, 053004 (2005).

\bibitem{Baker_2006}
Baker, S., Robinson, J., Haworth, C., Teng, H., Smith, R., Chirila, C., Lein,
  M., Tisch, J., and Marangos, J.
\newblock {\em {Science}}{ \bf {312}}, {424} ({2006}).,
Baker, S., Robinson, J.~S., Lein, M., Chirila, C.~C., Torres, R., Bandulet,
  H.~C., Comtois, D., Kieffer, J.~C., Villeneuve, D.~M., Tisch, J. W.~G., and
  Marangos, J.~P.\newblock {\em {Phys. Rev. Lett.}}{ \bf {101}} ({2008})

\bibitem{Farrell_2011}
Farrell, J.~P., Spector, L.~S., McFarland, B.~K., Bucksbaum, P.~H., G\"uhr, M.,
  Gaarde, M.~B., and Schafer, K.~J.
\newblock {\em Phys. Rev. A}{ \bf 83}, 023420 (2011).

\bibitem{wat:karl75}
Karlsson, L., Mattsson, L., Jadrny, R., Albridge, R.~G., Pinchas, S., Bergmark,
  T., and Siegbahn, K.
\newblock {\em J.\,Chem.\,Phys.}{ \bf 62}, 4745 (1975).

\bibitem{wat:reut86}
Reutt, J.~E., Wang, L.~S., Lee, Y.~T., and Shirley, D.~A.
\newblock {\em J.\,Chem.\,Phys.}{ \bf 85}, 6928 (1986).

\bibitem{Ning_2008}
Ning, C.~G., Hajgato, B., Huang, Y.~R., Zhang, S.~F., Liu, K., Luo, Z.~H.,
  Knippenberg, S., Deng, J.~K., and Deleuze, M.~S.
\newblock {\em {Chem. Phys.}}{ \bf {343}}, {19} ({2008}).

\bibitem{sfm:falg10}
Falge, M., Engel, V., and Lein, M.
\newblock {\em Phys.\,Rev.\,A}{ \bf 81}, 023412 (2010).

\bibitem{Smirnova_2009}
Smirnova, O., Patchkovskii, S., Mairesse, Y., Dudovich, N., Villeneuve, D.,
  Corkum, P., and Ivanov, M.~Y.
\newblock {\em Phys. Rev. Lett.}{ \bf 102}, 063601 (2009).

\bibitem{DiChiara_2009}
DiChiara, A.~D., Sistrunk, E., Miller, T.~A., Agostini, P., and DiMauro, L.~F.
\newblock {\em {Opt. Express}}{ \bf {17}}, {20959} ({2009}).

\bibitem{Wong_2010}
Wong, M. C.~H., Brichta, J.~P., and Bhardwaj, V.~R.
\newblock {\em {Opt. Lett.}}{ \bf {35}}, {1947} ({2010}).

\bibitem{Salieres_1995}
Sali\`eres, P., L'Huillier, A., and Lewenstein, M.
\newblock {\em Phys. Rev. Lett.}{ \bf 74}, 3776 (1995).

\bibitem{sfm:awas08}
Awasthi, M., Vanne, Y.~V., Saenz, A., Castro, A., and Decleva, P.
\newblock {\em Phys.\,Rev.\,A}{ \bf 77}, 063403 (2008).

\bibitem{sfm:petr10}
Petretti, S., Vanne, Y.~V., Saenz, A., Castro, A., and Decleva, P.
\newblock {\em Phys.\,Rev.\,Lett.}{ \bf 104}, 223001 (2010).

\bibitem{gen:vanl94}
van Leeuwen, R. and Baerends, E.~J.
\newblock {\em Phys.\,Rev.\,A}{ \bf 49}, 2421 (1994).

\bibitem{bsp:toff02}
Toffoli, D., Stener, M., Fronzoni, G., and Decleva, P.
\newblock {\em Chem.\,Phys.}{ \bf 276}, 25 (2002).

\bibitem{gen:herz66}
Herzberg, G.
\newblock {\em Molecular Spectra and Molecular Structure III: Electronic
  spectra and electronic structure of polyatomic molecules}.
\newblock Van Nostrand, New York,  (1966).

\bibitem{sfm:patc09}
Patchkovskii, S.
\newblock {\em Phys.\,Rev.\,Lett.}{ \bf 102}, 253602 (2009).

\bibitem{Saenz_2000}
Saenz, A.
\newblock {\em {J. Phys. B: At. Mol. Opt. Phys.}}{ \bf {33}}, {4365} ({2000}).

\end{thebibliography}
\end{document}